\documentclass[11pt,a4paper]{article}
\usepackage[T1]{fontenc}
\usepackage[utf8]{inputenc}
\usepackage{authblk}
\usepackage[square,numbers]{natbib}
\usepackage{hyperref}
\usepackage{graphicx}
\usepackage{wrapfig}

\usepackage[margin=1.0in]{geometry}

\usepackage{fancyhdr}

\fancypagestyle{plain}{%
  \fancyhf{}%
  \fancyfoot[C]{Copyright © 2022 for this paper by its authors. Use permitted under Creative Commons License Attribution 4.0 International (CC BY 4.0).\\\textit{Proceedings of the Conference on Applied Machine Learning for Information Security}, 2022}%
}

\usepackage{lipsum}

\title{Half-Day Vulnerabilities: A study of the First Days of CVE Entries}
\author[1]{Kobra  Khanmohammadi\thanks{kobrakhanmohammadi@geotab.com}}
\author[2]{Raphaël Khoury\thanks{raphaël.khoury@uqo.ca}}
\affil[1]{ Geotab Inc, Oakville, Ontario, Canada}
\affil[2]{Université du Québec en Outaouais,  Department of Computer Science and Engineering, Gatineau, Québec, Canada }
\date{}

\begin{document}

\maketitle

\begin{abstract}
The  National Vulnerability Disclosure Database is an invaluable source of information for security professionals and researchers. However, in some cases, a vulnerability report is initially published with incomplete information, a situation that complicates incident response and mitigation. In this paper, we perform an empirical study of vulnerabilities that are initially submitted with an incomplete report, and present key findings related to their frequency, nature, and the time needed to update them. We further present a novel ticketing process that is tailored to addressing the problems related to such vulnerabilities and demonstrate the use of this system with a real-life use case.  
\end{abstract}

\section{Introduction}

The  National Vulnerability Disclosure Database \cite{NIST} is the U.S. government’s repository of  vulnerability management data.  As presented in \cite{NIST}, the NVD defines a vulnerability as:
``A weakness in the computational logic (e.g., code) found in software and hardware components that, when exploited, results in a negative impact to confidentiality, integrity, or availability. Mitigation of the vulnerabilities in this context typically involves coding changes, but could also include specification changes or even specification deprecation (e.g., removal of affected protocols or functionality in their entirety).''

For each vulnerability, the NVD contains an entry, called a Common Vulnerability Enumeration (CVE), which records all relevant information about the vulnerability in a standardized manner. Amongst other information, the NVD contains a brief description of the vulnerability, a severity score, mitigation procedures, and a list of affected products and vendors, as well as a unique identifier. This information allows information technology professionals to rapidly identify, prioritize and patch vulnerabilities in the system they manage. 

Unfortunately, it is not uncommon for a CVE to be initially published with all or part of this information missing. Often, the report will be updated in the hours and days that follow its initial publication, and any missing section will be added to the CVE report, but this is not always the case. 

Incomplete CVE reports can have negative consequences on the security of information systems. Notably, the absence of a severity score makes it difficult to prioritize vulnerabilities, while the absence of a list of affected products makes it difficult for security managers to determine if they are exposed to a security risk. Most consequentially, the absence of mitigation forces them to weigh a difficult trade-off between exposing their firm to security risks and foregoing use of a software system.

In this paper, we examine how CVE reports are modified and updated in the first days after their initial disclosure. We make three main contributions:

First, we perform an empirical study, answering 7 research questions related to the vulnerability disclosure, thus shedding a light on the topic.  Second, we propose a novel ticking system that aids security professionals to perform vulnerability management in the presence of incomplete CVE reports.  Finally, we present we real-life use–case of our ticketing system, which we implemented at a large software firm. 

The remainder of this paper is organized as follows. Section \ref{sec:back} presents some background information. Section \ref{sec:setup} describes and motivates the setup of  our study. Section \ref{sec:emp} provides the results of the empirical part of our study. Our novel ticketing system is explained in Section \ref{sec:ticket} and a use-case is provided in Section \ref{sec:case}. Related works are given in Section \ref{sec:related}. Concluding remarks are given in Section \ref{sec:conclu}. 

\section{Background}\label{sec:back}
 
The  National Vulnerability Disclosure Database \cite{NIST} is the U.S. government’s repository of  vulnerability management data. Each vulnerability in the NVD is assigned a unique CVE identifier.  This database is an invaluable source of information  for security professionals since  few organizations have enough resources to research and find the vulnerabilities in every software asset that they rely upon. It is updated every two hours. 

For each vulnerability, NVD provides a score, by way of the Common Vulnerability Scoring System (CVSS). This score records a number of metrics about the vulnerability, most notably the  ‘Base score’ which represents the intrinsic characteristics of each vulnerability that are constant over time and across user environments. The Base Score is calculated based on two sets of metrics: the Exploitability metrics and the Impact metrics. The Exploitability metric represents the ease and technical means by which the vulnerability can be exploited and includes ‘Attack vector’, ‘Attack complexity’, ‘Privilege required’, ‘User interaction’ and ‘Scope’. The Impact metrics represent the direct consequence of a successful exploit and includes: ‘Confidentiality impact’, ‘Integrity impact’ and ‘Availability impact’. More details on the metrics are available in \cite{CVSS-V3}. 

The NVD provides two versions of CVSS (v.2 and v.3). Version 3 was released in 2015, and v.2 is no longer supported for new vulnerabilities. In this paper, we focus on the more recent v.3. The NVD calculates a quantitative value between 0-10 for CVSS v.3 base score. It also provides a qualitative ‘severity’ rankings of either "Low" (for base score between 0.1-3.9), "Medium" (for base score 4.0-6.9), "High" (for base score 7.0-8.9), or "Critical" (for base score 9.0-10).

Apart from vulnerabilities, NVD provides a list of software products for which a CPE (Common Platform Enumeration) label has been assigned. The CPE Dictionary is hosted and maintained at NIST and is available to the public. The CPE is a structured naming scheme for information technology systems, software, and packages. CPE provides a unique name for each product and version. We can identify a product by the name, vendor and version of the product shown in the CPE. A complete NVD  vulnerability report contains a list of CPEs showing the products containing such vulnerabilities. Unfortunately, as mentioned above, the NVD contains incomplete reports, and this information is sometimes missing. 

\section{Study Design and Motivation} \label{sec:setup}

We downloaded the NVD vulnerability datasets\footnote{https://nvd.nist.gov/vuln/data-feeds} every day for a period of three months from June 2021 to August 2021. The downloads were performed at midnight. During this period, the NVD published 40,813 vulnerability reports, covering 14,896 distinct CVEs with a unique ID. The NVD thus published  25,917  updates to vulnerabilities that already had been published during the period of the study. 

Some entries in our dataset are updates to CVE reports that were initially published before the onset of our study in June 2021. For such report, we were able to obtain the initial date of publication by referring ourselves to the "Published Date" field present in each report. This was the case for 846 entries in our dataset. However, 403 of these 846 entries were updates of much older reports (sometimes several years old), which include a v.2 CVSS score, but not the v.3 CVSS score.  We have opted to elide these reports from our study. 


For those vulnerabilities that were updated, the average number of updates is 2.74.  However, the number of updates is highly variable with some reports being updated as many as 17 times. 

This dataset forms the basis of our analysis, which seeks to determine how the information contained in a CVE entry changes during the first days after disclosure.  In particular, not all reported vulnerabilities initially have a complete report including its CVSS score, CPE list and mitigation resources. The NVD often reports a vulnerability soon after it's discovered and updates its report at a later date.  Therefore, having daily updates of the vulnerability  for a period of time allows us to study how frequently they are updated. 

More specifically, we attempted to answer the following research questions:

\begin{description}
    \item[RQ1] \textit{How many  vulnerabilities are initially reported without a CVSS score each day?}
    
    If a CVE entry does not contain a CVSS base score, it falls on the IT team in each company that is running the affected software to estimate key attributes of the vulnerably such as the ease of exploit and the potential impact. These attributes in turn affect the risk incurred by the vulnerability, and determine the priority of treating this vulnerability. The absence of a CVSS base score is thus a  problematic issue. 
    
    \item[RQ2] \textit{How long after the CVE is initially published until the CVSS score is finally reported? }If the CVSS score is routinely added shortly after the initial divulgation of the vulnerability, the problems associated with its initial absence are somewhat mitigated, and the security professionals in charge of taking corrective action can simply wait for the update that will contain the required information. 
   
   \item[RQ3]\textit{How many vulnerabilities (CVEs) are not initially assigned a CPE list?}
   Likewise, the absence of a CPE list hinders the ability of IT professionals to patch systems affected by the vulnerability and take other actions to prevent exploitation, since it makes it difficult to identify the organizational assets that are affected by the underlying vulnerability. 
   
    \item[RQ4]\textit{How long after the CVE is initially published until the related CPE is finally reported? }
    As is the case for the CVSS, the absence of CPE in a vulnerability report is specially problematic if the vulnerability is not updated to include this information shortly after its initial publication. 
    
    \item[RQ5]\textit{ How many  vulnerabilities have no proposed mitigation approaches, including update or workaround?  }
    If a vulnerability is reported without a proposed mitigation, it forces IT professionals in any company that runs the affected software into a difficult calculus between exposing themselves to a possible attack, or foregoing the use of the software. 
    
    \item[RQ6]\textit{Are there vendors (CPE) that are more likely to report a vulnerability without a CVSS rating and\textbackslash or a mitigation?} 
    Vendors that consistently report complete vulnerability reports in a timely manner can be thought of as providing  an added value to their users.
    
    \item[RQ7]\textit{Is there a statistically significant difference  in the CVSS scores of vulnerabilities that are initially reported without a CVSS score and those that are? }
    If vulnerabilities that are initially reported without a CVSS score turn out to be high severity vulnerabilities, then it may be appropriate for the prudent security professional to prioritize such vulnerabilities, alongside with those that are known to be high-risk.

 \end{description}

The python scripts used to perform the statistical analysis are available on the author's repository\footnote{https://github.com/kkhanmohammadi/nvd\_cve\_study}. 

\section{Empirical findings}\label{sec:emp}

\subsection{RQ1:  How many  vulnerabilities are initially reported without a CVSS  base score each day?}

Some vulnerabilities are initially reported with no severity score assigned to them. The main reason for this situation is that these vulnerabilities have not yet been completely investigated because of the time constraints. Usually, a CVSS rating will be assigned to the vulnerability a few days later. When a vulnerability is reported with no CVSS score, security analysts from each company that is running the affected code must conduct a manual investigation in order to determine what remedial steps must be taken, and to assess the nature and urgency of the vulnerability. In this case, the severity of the vulnerability can be determined by what informational asset the vulnerability relates to, how central that asset is to the organization, and by the nature of the vulnerability.

Figure \ref{no-severity} shows the distribution of the number of vulnerabilities initially reported  without a CVSS score for the period of our study. This provides an estimate of how many such vulnerabilities one might expect to encounter daily. We found that  11 473 out of 40 813 (28\%) vulnerability reports published during three months of study  had no assigned CVSS base score. These reports represent 5270 out of 14 896 (35\%) distinct vulnerabilities. The average  number of vulnerabilities reported with no CVSS base score each day is 139.9.


\subsection{RQ2: How long after the CVE is initially published until the CVSS score is finally reported?}
\begin{figure}[!tb]
  \centering
  \begin{minipage}[b]{0.4\textwidth}
    \includegraphics[width=\textwidth]{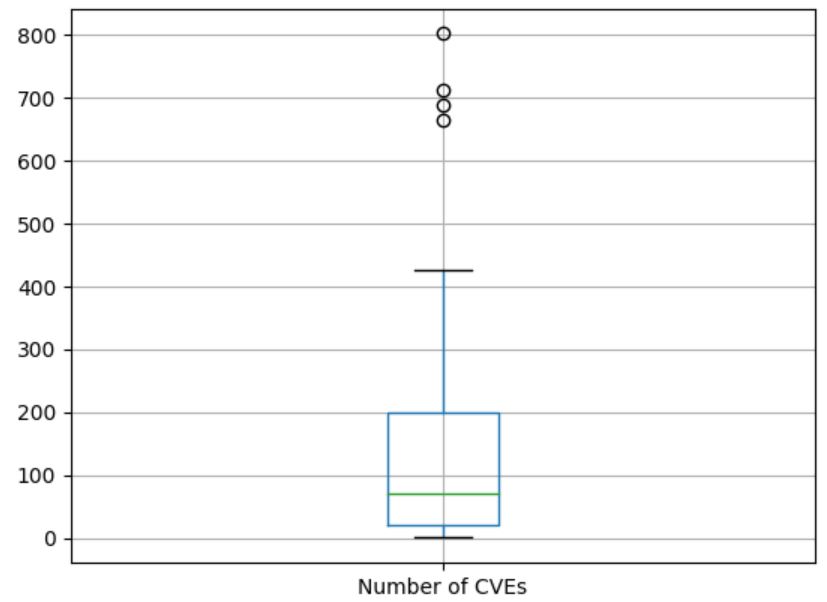}
    \caption{Distribution of number of vulnerabilities with no assigned base score per day.}
    \label{no-severity}
  \end{minipage}
  \hfill
  \begin{minipage}[b]{0.4\textwidth}
    \includegraphics[width=\textwidth]{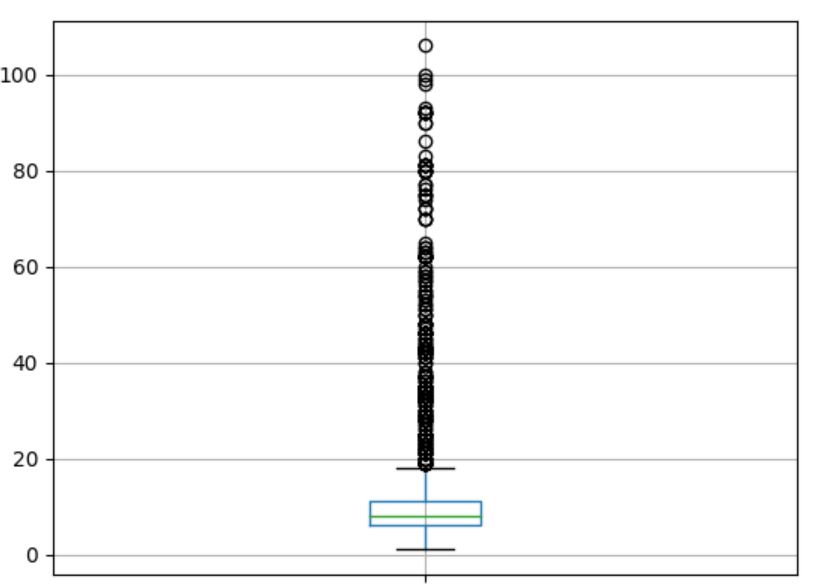}
    \caption{Number of days between initial report of a vulnerability and the inclusion of a CVSS score.}
    \label{date_differences}
  \end{minipage}
\end{figure}

As it mentioned above, the absence of a CVSS score is somewhat mitigate if the CVE report is rapidly updated with the missing information. Figure \ref{date_differences} shows the distribution of the number of days that elapse between the initial date of reporting of a vulnerability and and the date on which it is updated with the inclusion of a CVSS score. Vulnerabilities that are initially reported with a CVSS score are naturally omitted from this statistic. We also omitted any vulnerability which was initially introduced without a CVSS score and for which a score had not yet been provided by the end of the period covered by our study. 

 
As mentioned above, our dataset contains 5270 CVE entries for which no CVSS score was initially provided. Out of these 5270 entries, 3612 (69\%) were eventually  updated with a CVSS v.3 base score. An additional 334 entries (6\%) did receive an update, but were not assigned a CVSS v.3 score as part of that update. Finally,  1324  (25\%) were never updated for the duration of our study. The fact that some of these entries may eventually have been assigned a CVSS v.3 score at a moment that falls outside of the time frame of our study is a threat to the validly of our results. 

As can be seen from the Figure \ref{date_differences}, the average number of days until these entries are updated with a CVSS score  is 11.62 days. 

\begin{figure}[!tb]
  \centering
  \begin{minipage}[b]{0.4\textwidth}
    \includegraphics[width=\textwidth]{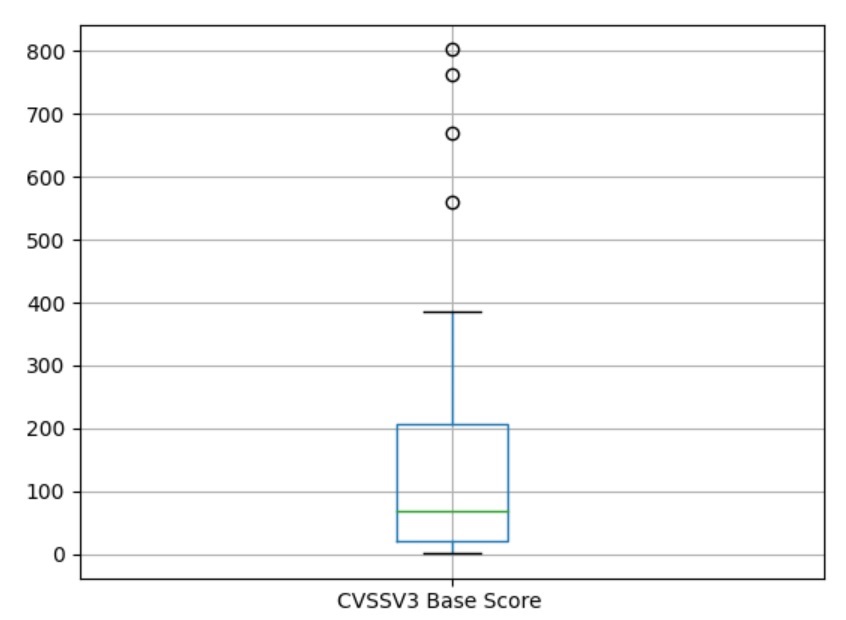}
    \caption{Distribution of number of vulnerabilities initially reported with no CPE.}
     \label{no-cpe}
  \end{minipage}
  \hfill
  \begin{minipage}[b]{0.4\textwidth}
    \includegraphics[width=\textwidth]{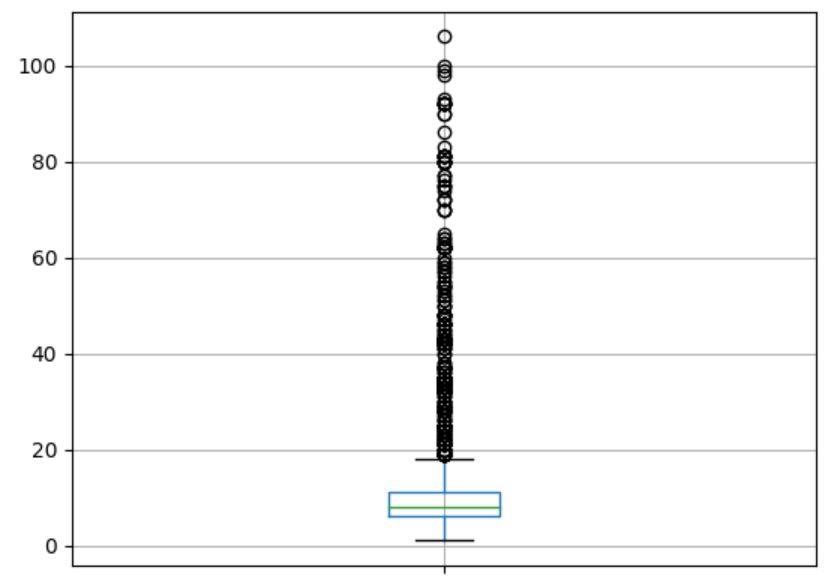}
    \caption{Number of days  between  the initial report and first updated report containing CPEs.}
    \label{date-difference-cpe}
  \end{minipage}
\end{figure}


\vspace{1cm}
\subsection{RQ3: How many vulnerabilities (CVEs) are not initially assigned a CPE list?}
Vulnerabilities are also sometimes initially reported without a list of vulnerable products (CPE).  This makes it much more difficult to identify the organizational assets that are affected by the vulnerability in question.  During the period of our study, 7748 out of 14,896 (52\%) vulnerabilities were initially reported without a CPE list. Of these 7748,  2248 (29\%) were eventually updated with the inclusion of a CPE list during the three month of our study. When considering reports, rather than individual vulnerabilities, we find that 10965 out of 40813 reports  (27\%)  did not contain a CPE list.  As shown in Figure \ref{no-cpe}, the average  number of vulnerabilities without CPEs reported each day is 133.7.


\subsection{RQ4: How long after the CVE is initially published until the related CPE list is finally reported?}

The distribution of the number of days that elapsed between the initial report of a vulnerability which has no CPE list included,  and the first  update to this report that assigns it a CPE list is shown in Figure \ref{date-difference-cpe}. The average is 11.5 days. This is a considerable amount of time, and indicates that it would be imprudent for security professionals to wait until a CVE is updated with its CPE list before making a determination as to whether or not they are exposed to the underlying vulnerability. We will return to the problem of security management in the absence of CPEs in the next section. 

As mentioned above, there were 5128 vulnerabilities with no CPE list during the 3 months of our study. Among them, 2649  (51.65\%), were eventually updated with the inclusion of a CPE during the three months of our study. It is also interesting to note that an additional 270 (5\%)  vulnerabilities did received an update, but that this update did not include the missing CPE. This indicates that providing a CPE is not always the overarching concern of the security professional that discover and maintain these vulnerabilities. 

%

\subsection{RQ5: How many  vulnerabilities have no proposed mitigation approaches, including update or workaround?}

\begin{figure}[!tbp]
  \centering
  \begin{minipage}[b]{0.4\textwidth}
    \includegraphics[width=\textwidth]{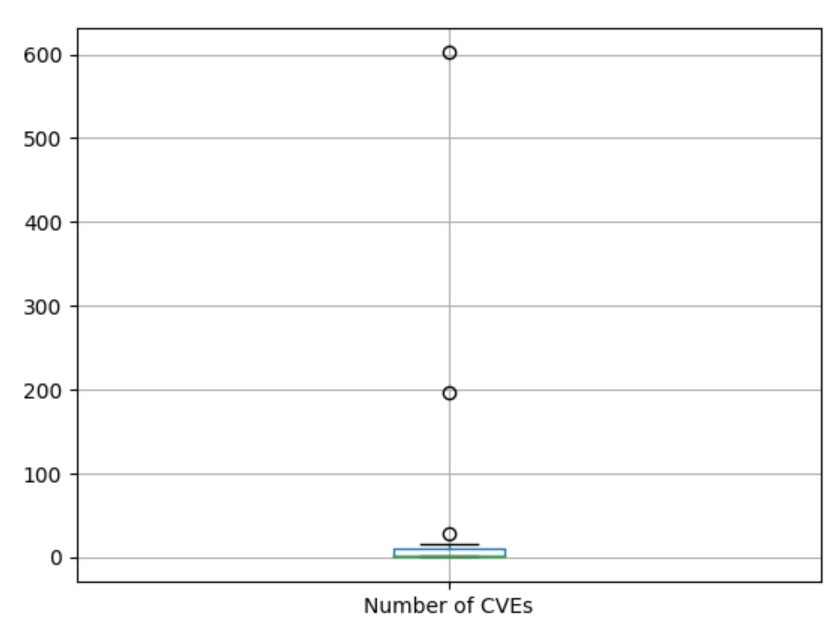}
    \caption{Distribution of number of vulnerabilities initially reported with no mitigation.}
    \label{no-references}
  \end{minipage}
  \hfill
  \begin{minipage}[b]{0.4\textwidth}
    \includegraphics[width=\textwidth]{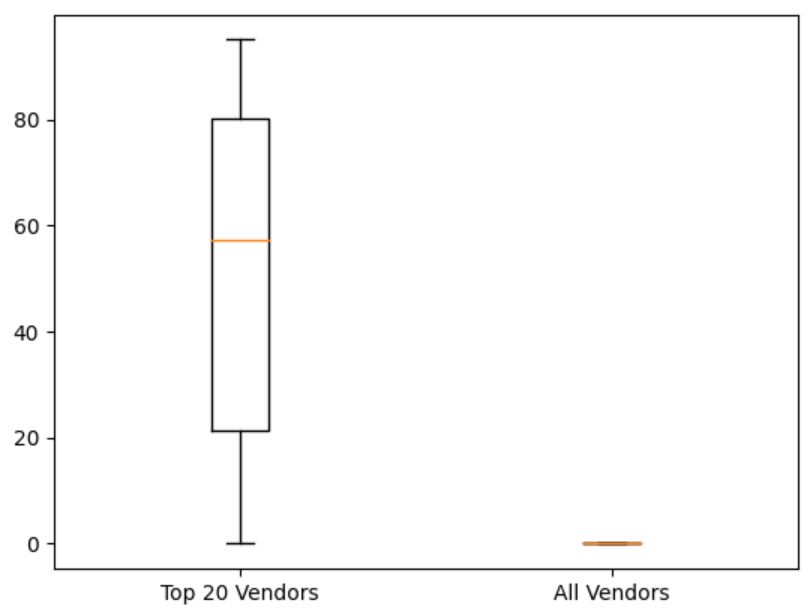}
    \caption{Percentage of vulnerabilities with no CVSS score in the initial report.}
    \label{vendors}
  \end{minipage}
\end{figure}

A CVE report contains a section titled "References to Advisories, Solutions, and Tools", which presents the method for mitigating the vulnerability. The proposed solution is usually updating the software to the latest version. This section of the CVE entry contains links to websites explaining the mitigation process. When the section is empty,  no  update or workaround for the vulnerability is available. Usually, the mitigation is included in the CVE entry simultaneously with the CVSS score. When  no mitigation approach is provided for a vulnerability, it falls to the  organization running the vulnerable code to make decision on whether or not to continue using the code in question.  Figure \ref{no-references} shows the distribution of vulnerabilities with no suggested mitigation. For the period of our study,  894 out of 40,813 (2\%) vulnerabilities were initially reported with no mitigation included in the report. When considering distinct vulnerabilities with unique CVE IDs, 381 out of 14896 (2\%) vulnerabilities fall in this category. The average number of vulnerabilities reported each day that lack this information is 47.05.

\subsection{RQ6: Are there manufacturers (CPE) that are more likely to report a vulnerability without a CVSS rating and\textbackslash or a mitigation?}
For each vulnerability, we extracted the name of the associated vendor or vendors as recorded in the CPE list.  In cases where the CVE entry did not initially contain a CPE list, we obtained this information from subsequent updates to the entry.  From this data, we identified the top 20 vendors with the highest percentage  of vulnerabilities initially reported with no CVSS score, as well as the top 20 vendors with the highest percentage of CVEs submitted with a CVSS score from the onset. These results are shown in Figure \ref{vendors}.

Across all vendors, the average percentage of vulnerabilities initially reported without a CVSS score is 35\%. This number jumps to 82.63\% for the top 20 vendors most likely to submit an incomplete vulnerability report. The bar chart in Figure \ref{vendors} depicts the distribution of the percentage of vulnerabilities with no CVSS base score for top 20 vendors most likely to submit such reports, in comparison to that of all vendors. This is a substantial difference, and one which we found to be statistically significant by performing  a Wilcoxon-Mann-Whitney test (p-value $\approx$ 0).

Anderson, in his seminal paper \cite{anderson}, argued that the inability of software vendors to provide  objective metrics about the quality of their code  to potential clients induces a "market for lemons", which favors lower quality products. This is because a client who is unable to evaluate the degree of security of a product is naturally unwilling to pay a premium for the benefit of a more secure product. Since the practice of consistently including a CVSS score and a mitigation in CVE reports offers tangible security benefits, it helps mitigate the problem identified by Anderson, and could potentially be a part of a strategy by a vendor who wishes to distinguish himself from his competitors by offering security guarantees about his product.


\begin{figure}
\centering
\begin{minipage}{.5\textwidth}
  \centering
  \includegraphics[width=1\textwidth]{ 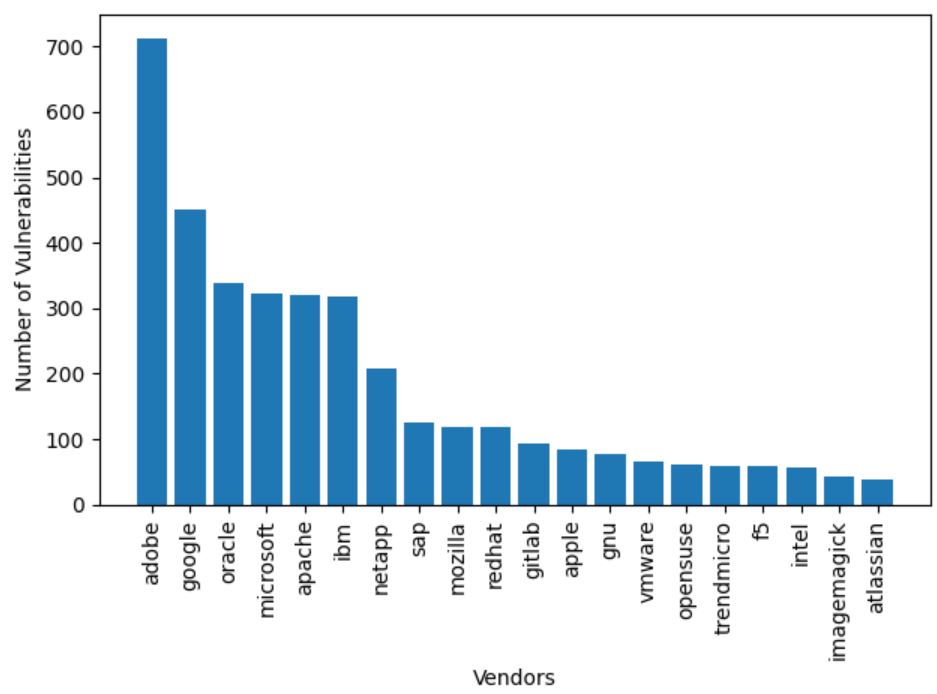} \\
  \label{vendors_initial_top20}   
\end{minipage}%
\begin{minipage}{.5\textwidth}
  \centering
  \includegraphics[width=1\linewidth]{ 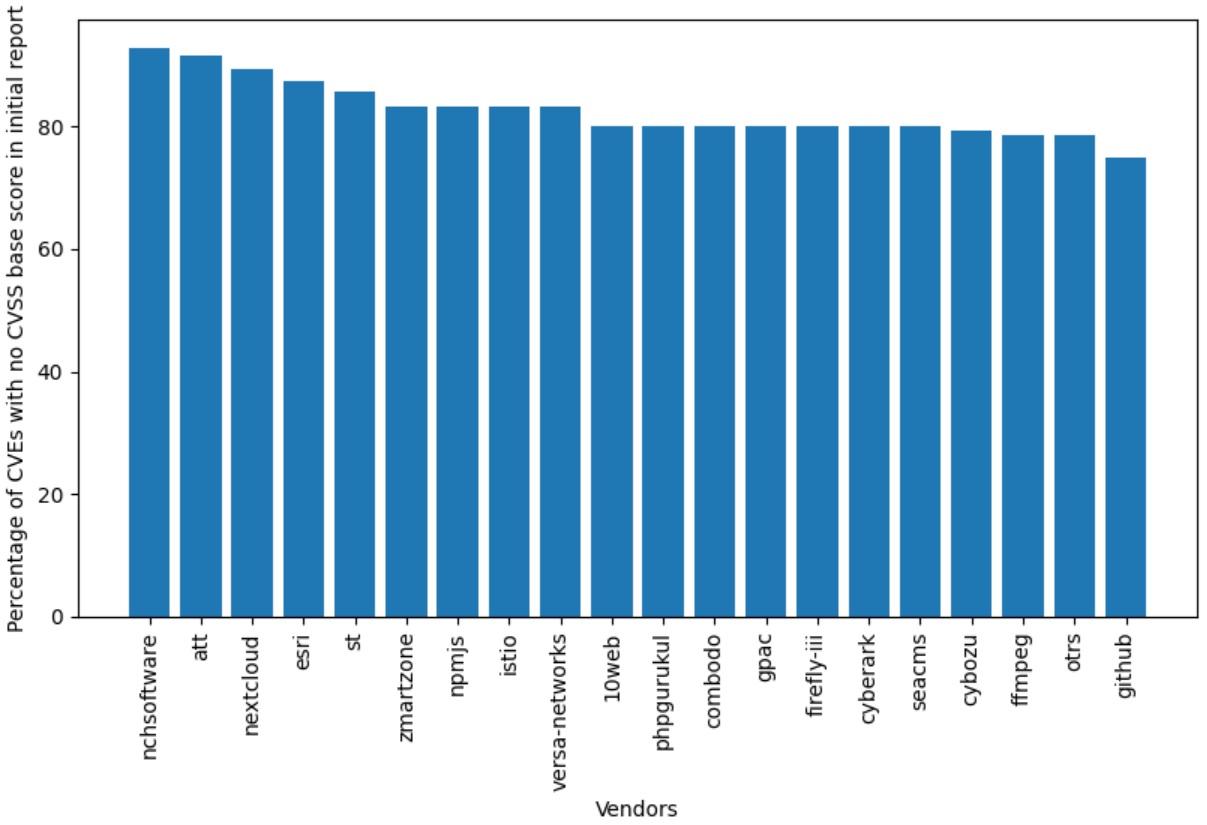}\\
  \label{vendors_updated_top20}
\end{minipage}
\caption{Top 20 vendors most likely to submit an initial vulnerability report with a CVSS score (left) and without a CVSS score (right).}
\label{vendors}
\end{figure}

\subsection{RQ7:Is there a statistically significant difference in  CVSS score values  between vulnerabilities that are initially reported without a CVSS score and those that are?}

Another important question is to determine if vulnerabilities  for which a CVSS score is only provided later have a different distribution of CVSS score values in comparison to vulnerabilities containing a CVSS score in their initial report. If such vulnerabilities were found to be likely to be high severity, then security professionals would be justified in prioritizing them even though their severity score is not known, alongside with those vulnerabilities that are known to be high-risk.   

Table \ref{table:scoreinitial} shows the percentage of vulnerabilities with a CVSS score in their initial report and those for which a CVSS score is later provided in an update. We performed a Wilcoxon-Mann-Whitney test, which  showed that there is no statistically significant difference between these two distributions of vulnerabilities scores (p-value is 0.44 which is greater than 0.05).

\begin{table}[]\begin{center}
\begin{tabular}{|l|c|c|}
\hline
CVE Base Score level & CVEs with  CVSS in  &  CVEs with CVSS  \\ 

  &   the initial report &    reported at a subsequent date \\ \hline
CRITICAL & 124(17\%) & 425(13\%) \\ \hline
HIGH     & 373(49\%) & 1434(45\%) \\ \hline
MEDIUM      & 232(30\%) & 1338(40\%) \\ \hline
LOW      & 28(4\%) & 81(2\%) \\ \hline
\end{tabular} \caption{Distribution of vulnerability score according to whether the score is initially present or not.}
\label{table:scoreinitial}
\end{center}
\end{table}


\subsection{Key Findings}

We found that it is surprisingly common for vulnerabilities to be initially published in the NVD database with key information missing from the report, notably the CVSS score (35\%), the CPE (52\%) and the mitigation (2\%).  In cases where the CVSS report is missing, the average number of days until its inclusion is 13.5 days. For CPE, the corresponding value is 14.5 days. Furthermore, as many as 35\% of vulnerabilities are never assigned a CPE. These numbers are vary widely from one vendor to another, a fact that more assiduous vendors might choose to capitalize on. 

Only about 2\% of vulnerabilities are not assigned a mitigation. Vulnerabilities that are initially published without a CVSS score do not seem to differ widely with respect to severity from those that do include the score from the onset. 

\section{CVE Matching System}\label{sec:ticket}

The results presented in the previous section show that incomplete CVE reports are common, and that this fact can hinder the process of promptly responding to security vulnerabilities.  This is particularly problematic since organizations are often required to implement an incident response  plan, both because of their commitment to specific SLA, and in order to maintain various security certificates such as ISO 27001. This plan requires them to mitigate any vulnerability reported in their software assets within a time period that varies according to the risk severity of the vulnerability.

In general, vulnerability management includes the following steps: identifying vulnerabilities on the organization's assets, measuring the threats they pose, estimating the associated risk level and finally mitigating the risk by applying solutions to resolve the vulnerabilities. The absence of a CPE list and of a CVSS score in a new CVE entry makes this process much more difficult. In this section,  we propose a methodology to use NVD's vulnerability dataset to identify the vulnerabilities that relate to an organization’s assets in a context where the CPE list may be missing from a CVE. This methodology also aids in the process of creating tickets. A ticket in a service desk platform is an event that must be investigated or a work item that must be addressed.

\begin{wrapfigure}{L}{0.5\textwidth}
\centering
\scalebox{1}{
  \frame{\includegraphics[width=\linewidth]{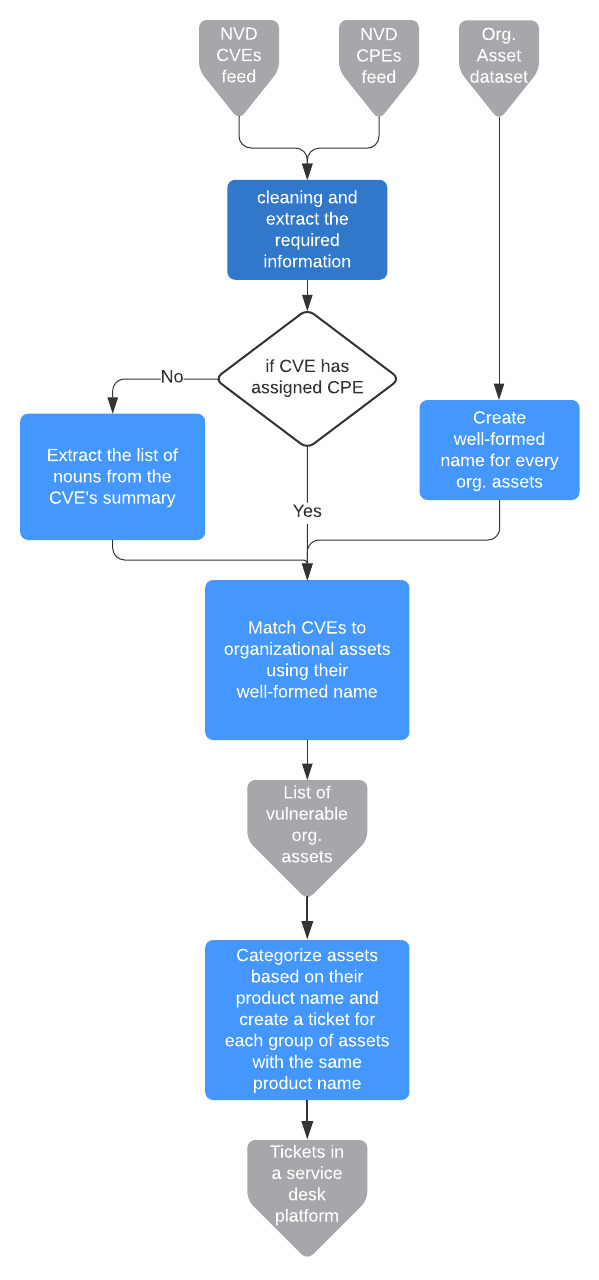}}}
  \caption{Methodology for relating CVEs to an organization assets in the absence of a CPE.}\label{plan}
\end{wrapfigure}

Figure \ref{plan} schematizes our proposed methodology. The inputs are, (1) the set of new vulnerabilities reported by NVD in the previous 24 hours, (we assume that the process of fixing vulnerabilities is performed daily); (2) the latest version of the CPE dictionary from the NVD  and; (3) the organization’s asset inventory.  Our methodology allows for the creation of tickets even in the absence of a CPE list in the CVE report, and further coalesces multiple vulnerabilities that target the same system in a single ticket, which aids in prioritizing and treating the vulnerability. 

The next step of the vulnerability mitigation process is to relate the CVEs to organizational assets. If the CVE contains a CPE list, and if the CPE label of every organizational asset is listed in the organization's asset list, then this is a straightforward process. However, as discussed above, the CPE list is often omitted from the CVE report. There may be a variety of reasons for this. Notably, not all products are assigned a  CPE label. Note that it is the responsibility of each organization to report new versions of their products to the NVD, so that a new CPE labels can be  issued, and this process is not always performed promptly. Moreover, an organization's asset inventory may not be complete. For example, a security manager may overlook a vulnerability if the organization's asset inventory fails to record the version of the software under threat, leading him to skip over some CVE reports he wrongly sees as unrelated to his organization's assets. Thus, the CPE label of a vulnerable software may be missing from either the vulnerability report, the asset list, or both. 

Consequently, it is not always possible to rely on the CPE label reported in a CVE report to determine to which assets in an organization are related to a given vulnerability.  In our methodology,  we introduce the notion of the \textit {well-formed name} of a product.  A product's well-formed name is a canonical string that contains the product's name, vendor and version, in a dictionary format (\{name:product's name, vendor:vendor's name, version:product's version\}). The well-formed name can serve as an alternative canonical representation for a product in an organization's asset list when the CPE label is missing. For each organizational asset in the organization, we first check if there exists a CPE in the NVD's CPE dictionary (https://nvd.nist.gov/products/cpe). If so, the product's well-formed name consists of the product's name, vendor and version, as recorded in the CPE dictionary. If no CPE is found for an asset, we  manually construct a well-formed name containing  the name, vendor, and version of the asset.

In some cases, the asset inventory does not have a canonical format for recording the assets name in a uniform manner, thus necessitating an additional standardization (cleaning) phase. Standardizing the asset's name provides better matching between the name used by the organization  and the product names recorded in the NVD's vulnerability report. Some possible standardizing methods include deleting any information written in parentheses or curvy brackets, deleting numbers and dates, and  deleting very common names in assets such as “system”, “software”, “library”, "version" or “app”. For example, for the product listed with the product name: 'R2D2 Beta version 3.0.1.16' and vendor name: 'Geotab Inc.' in an asset inventory, the well-formed name is $\{name:'r2d2', vendor:'geotab', version:'3.0.1.16'\}$.

The absence of a CPE list in the vulnerability report also introduces similar difficulties.  As  shown in Figure \ref{plan}, if the CVE entry includes a CPE list, we can simply use it to derive the well-formed name of the vulnerable products. However, if there is no CPE list, we attempt to extract the name of the vulnerable product from the summary present in the CVE report. The name of the product is a noun that normally appear somewhere in the summary of a CVE, so the main challenge is to identify the name in the summary.  To this end,  we first use the NLP library Stanza \footnote {https://stanfordnlp.github.io/stanza/} to extract a list of nouns from the summary section of a CVE report. In what follows, we refer to this list as the  "summary-nouns" list.  In the next step, for each vulnerability, if a CPE list is present in the CVE, we check the organization’s list of well-formed names to determine if the organization runs this software as part of its information assets. If the CVE report does not include a CPE list, we check if the "summary-nouns" of  the CVE contains the name of any of the  organization’s assets. 

Attempting to identify assets related to a vulnerability  based on the nouns in the CVE'S summary  will cause some false positives to occur. This is because different products, by different vendors may have similar or partially similar names (for eg.  VirtualBox and Box). Furthermore, the  summary likely contains a number of nouns other than the name of the product. Some of these nouns may coincide with the names of products by other vendors. For example, a summary may explain that the vulnerability is  of type “SQL injection”. Here, “SQL” will be identified  as a noun and may be cause a false positive match with a product called “SQL server”.

Moreover, we find out that some of the names of software are common, short (1-2 letter) words, which leads to false positives in when matching CVEs to an organization's asset list. Therefore, we first applied a filter that eliminates such common words from the CVEs’s summaries. This filter was constructed as follows: First, we extracted the list of all 2020 CVEs for which a CPE was provided, and created a list of all products as well as a list of all vendors that occurred in CVEs that year.  We also extracted a list of nouns that occurred in the summary descriptions of CVEs  for that year. We then compiled two lists of vendor names and product names respectively that appear in the description of a CVE, but not in that CVE’s CPE list.   Such words are likely to trigger false positive, but only if the related product or vendor name appears in the enterprise’s asset list. This list,  as well as the code required for cleaning and matching of product names to summary of CVEs, are available on the author’s repository\footnote{github.com/kkhanmohammadi/nvd\_cve\_study}.

It is important to stress that identifying every  vulnerability related to the company's assets and reporting each of them in a separate ticket is not a adequate practice. Indeed, doing so would lead to a large number of tickets. However,  multiple vulnerabilities reported on the same day may relate to the  same software. Since the most common solution for mitigating a vulnerability is updating the software to the latest version, it makes sense to group CVE reports that relate to the same software in a single ticket.  This grouping is made irrespective of the version of the software, since the mitigation will likely involve applying an update. 


\section{Case study}\label{sec:case}
We implemented the approach proposed in Figure \ref{plan} in a branch of Geotab Inc.\footnote{www.geotab.com}, a company that  provides solutions for fleet management and vehicle tracking.

Table \ref{table:casestudy}, summarizes our use of the framework with respect to the vulnerabilities in Geotab's assets for a period of six months between December 2020 and May 2021. Since the list of assets changes daily, we show  the average for the number of assets during that period --- around 500k products. This includes every instance of every software asset utilized by the firm. We grouped the assets according to their names and vendors. In total there were 446 678 such groups. 

Each day, an average of 39 asset groups were identified as having at least one vulnerability. However, the average  number of vulnerable assets (without grouping)  is 163. On average, for those assets that present vulnerabilities, around 4.5 CVEs are related to that asset.

As explained in section \ref{sec:ticket}, we group the vulnerable assets according to their names without considering the version and report every vulnerability related to a group of assets with the same product name in a single ticket. Thus, a single ticket may refer to several vulnerabilities. Subsequently, these tickets  are be recorded in a vulnerability management software and  will subsequently be addressed by a security analyst. As shown in Table \ref{table:casestudy}, in our case study, on average, each day 7 tickets were issued that were related to vulnerabilities that did not have a severity rating at the moment of the creation of the ticket and our approach was able to match them correctly to the products in the company. 

As explained, we expected to get some false positives in reporting vulnerabilities not related to the company's assets because of our reliance on an NLP library to automatically extract product names from the ``Summary'' section of each vulnerability report in datasets. In our case study, on average, 5 tickets were false positives whose reported vulnerabilities were not related to Geotab assets. This number was judged by our partners at Geotab to be sufficiency small as to not outweigh the benefits of the proposes ticketing system. 
\begin{table}[]\begin{center}
    
\begin{tabular}{|l|l|} \hline
Average number of  assets including different products,
&  513 280  \\ 
vendors, and different version per day&
  (divided in 446 678 groups) \\ \hline

Average number of CVEs matched to assets  &  \\ 
 per day,  including CVEs with no specified & \\  
 CPE,  during 6 months, (Dec 2020-May 2021)                          & 39  \\ \hline
 
 Average number of CVEs with no specified CPE matched  &  \\ 
to assets per day, (Feb 2021) 
 & 33  \\ \hline
 
Average number of vulnerable assets  & \\  per day including assets related to CVEs & \\  with no specified CPE during 6 months & \\
(Dec 2020-May 2021)    & 163 \\ \hline

Average number of CVEs mapped to each asset records                                                                                                  & 4.5 \\ \hline

Average number of tickets per day, i.e.       & 11  \\ \hline

Average number of tickets per day with no specified CPE       & 7   \\ \hline

Average number of false alarm tickets                                                             & 5 \\  \hline
\end{tabular} \caption{Statistics on vulnerabilities related to assets in Geotab case study }
\label{table:casestudy}
\end{center}
\end{table}

\section{Related work}\label{sec:related}
Much of the literature on cybersecurity vulnerability management approaches the topic from the perspective of a specific industry. For example, \cite{SappalProwse}, \cite{1069} and \cite{chen2018inhospitable} focus on cybersecurity risk assessment scoring in the specific context of the heath industry. These studies develop  cybersecurity vulnerability management system that emulate the existing practices in maintenance of medical systems for responding to the challenges of managing cybersecurity vulnerabilities.

Likewise, Mantha and De Soto \cite{construction} proposed an approach that customizes the CVSS scoring system for the needs of construction projects while Tang et al. \cite{7968482} studied challenges in risk assessment of big data systems. Janiszewski et al. \cite{1391} proposed an approach for performing risk assessment at the national level, where a large number of institutions must be considered. The main challenge of risk assessment at the national level is the heterogeneity of institutions (and sectors) which complicate the risk estimation process. They present a novel quantitative risk assessment  and carry out risk estimation in real time. In their proposed approach, they identify institutions' services and estimate the risk based on the criticality of services and the criticality of relationships between each service. Haastrecht et al. \cite{1145} address similar challenges for small and medium size enterprises and outline the data requirements that facilitate automating risk assessment.

A number of papers focus on the risk assessment part of vulnerability management. Those papers mostly suggest  approaches to quantify risks associated with vulnerabilities.
Wang et al. \cite{Wang2020ABN} propose a novel approach for cybersecurity risk assessment. The approach uses a Bayesian network to improve the statistical distributions that can be used to estimate cybersecurity risks and also to improve the extensibility of the taxonomy model used to classify cybersecurity risks into a set of quantifiable risk factors. Zhang et al. \cite{quteprints} proposed an approach that uses fuzzy probability in a Bayesian network for predicting the propagation of cybersecurity risks. King et al. \cite{King} characterize human factors as a contribution to cybersecurity risk. Allodi et al. \cite{Allodi} present a model that leverages  the large amount of historical data available from the IT infrastructure of an organization’s security operation center to quantitatively estimate the probability of attack. Fielder et al. \cite{fielder2018risk} studied how uncertainties in risk assessment affect cybersecurity investments. They utilize a game-theoretic model to derive the defending strategies even when knowledge regarding risk assessment values is not accurate.

The automation of risk assessment is also the topic of active research.  Kasprzyk et al. \cite{Kasprzyk2016} propose an approach for automating risk assessment for IT systems. They present adjustable security checklists and standardized dictionaries of security vulnerabilities and vulnerability scoring methods. Syed \cite{article} proposed an approach for a Cyber Intelligence Alert (CIA) system that issues cyber alerts about vulnerabilities and countermeasures.

Sabillon at al. \cite{Sabillon2017ACC} reviewed the best practices and methodologies of global leaders in the cybersecurity audit arena and presented their scope, strengths and weaknesses. They also proposed a comprehensive cybersecurity audit model to be utilized for conducting cybersecurity audits in organizations and governmental institutions.  Roldán-Molina et al. \cite{ROLDANMOLINA2017568} studied commercially available tools that can be used to perform risk assessment and decision making in the cybersecurity domain. They analyzed their properties, metrics and strategies and assessed their support for cybersecurity risk analysis, decision-making and prevention for the protection of an organization's information assets.

A number of researchers have mined the NVD for actionable information of vulnerabilities and threats, a line of research in which this study places itself.

Khoury et al. \cite{cveInIoT} compared the studies the CVSS scores of  vulnerabilities exploited by IoT botnets and found that they differ substantially that remain unexploited by adversaries. Murtaza et al. \cite{w8} conducted an empirical study of the NVD  to detect trends of changes in software  vulnerabilities over six years. They used NVD as their main source of data to mine six years of software vulnerabilities,  from  2009 to 2014 and were able to predict the characteristics of future vulnerabilities in code, based on previous ones.

Na et al. \cite{w9} proposed a classification method for categorizing CVE entries into vulnerability type using naïve Bayes classifiers. Neuhaus  et al. \cite{w11}  tackled the same task, using Latent Dirichlet Allocation (LDA). Frei et al. \cite{w10} studied the delays between the time a vulnerability is disclosed in the NVD and the time a patch is published. They found that software vendors are slow to provide patches despite the fact that attacks that exploit zero-day vulnerabilities are an increasing concern. 

\section{Conclusion}\label{sec:conclu}
In this paper, we performed a empirical study of  vulnerabilities that are initially submitted with an incomplete report. We found that such reports are common, and that considerable time may elapse before they are  updated. Consequently,  we propose a novel ticketing system that aids in vulnerability management in the presence of incomplete vulnerability reports. Finally, we demonstrate the use of this system with a real-life use case.

Further research is needed to aid security professionals in dealing with incomplete reports. This paper lays the foundation by creating a ticketing systems that sidesteps problems associated with a missing CPE list. In the future, we would like to incorporate functionalities that predict the severity and ease of exploitation of the vulnerability if these datum are absent from the CVE report--- a common occurrence according to our results in Section \ref{sec:emp}. The task of the security analysts would also benefit from an automatic mechanism to detect duplicate CVE entries.

\bibliography{citations}

\end{document}